\RequirePackage[l2tabu, orthodox]{nag}
\documentclass{stsci_report}
\usepackage[pdftex,
pdfauthor={A. M. Guzman},
pdftitle={Analysis of Sink Pixels in ACS/WFC},
pdfsubject={},
pdfkeywords={ACS, CCD, Hubble Space Telescope, HST, Space Telescope Science Institute, STScI, Advanced Camera for Surveys, Wide Field Channel, Sink Pixels, columns, WFC, readout, amplifiers}]{hyperref}
\usepackage[section]{placeins}
\usepackage{booktabs}
\usepackage{rotating}
\usepackage{natbib}
\usepackage{url}
\usepackage{graphicx}
\usepackage[font=footnotesize,labelfont=bf]{caption}
\usepackage{subcaption}
\usepackage{nameref}
\usepackage{hyperref} 
\usepackage{array} 
\usepackage{color}
\usepackage{amsmath}
\usepackage{amssymb}
\usepackage{gensymb}
\usepackage[bottom,symbol]{footmisc}

\usepackage{float}
\usepackage{aliascnt}
\newaliascnt{eqfloat}{equation}
\newfloat{eqfloat}{h}{eqflts}
\floatname{eqfloat}{Equation}
\usepackage{adjustbox}
\usepackage{threeparttable}
\usepackage{booktabs}

\copyrighttext{Copyright \copyright \the\year\ The Association of Universities for Research in Astronomy, Inc. All Rights Reserved.}

\title{\textbf{Evolution of Sink Pixels in ACS/WFC and Connection to Charge Transfer Efficiency}}
\presubtitle{\flushleft{\includegraphics[width=5cm]{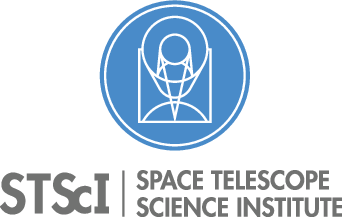}} \\ \hfill Instrument Science Report ACS 2024-01}
\author{A. M. Guzman, J. E. Ryon}
\date{April 29, 2024}

\begin{document}

\maketitle

\abstract{In our study spanning 2015-2021, we examined sink pixels (SPs) in the Advanced Camera for Surveys Wide Field Channel (ACS/WFC) using dark and SP reference files. SPs are pixels with values $\le$ $-10$ electrons below the local background of LED-flashed short (0.5 sec) darks, that collect and trap significant charge during readout. Analyzing seven years of short dark data, we assessed SP creation and persistence. In this time frame, 5,430 SPs were created in WFC1 and 5,649 SPs in WFC2, with creation rates of about 2.15 pixels/day and 2.23 pixels/day, respectively. These calculations allowed us to detect 44,068 SPs, not including SP trails, in the detector by the end of 2021, constituting approximately 0.25\% of the science frame. We found it is rare for SPs to return to a typical, non-negative pixel value. We observed more flagged SPs near the serial register than the chip gap. Skewed histograms for the $y$-position distribution, exhibiting a local peak in the distribution of SPs very near the chip gap described as the ``bounce-back" effect, were evident for both WFC1 and WFC2, while the $x$-position distribution remained uniform. Examining CTE-corrected images from 2015, 2018, and 2021 revealed consistent trends, with the gradient getting steeper over time due to CTE losses, which is also worse for pixels further from the serial register. We simulated the CTE-impacted readout of a short dark exposure with uniformly distributed SPs, to assess how CTE influences SP detectability.  While the gradient effect was reproduced, the local peak near the chip gap was not. Filling in of SPs by CTE charge-release during readout appears to explain most of the gradient in the $y$-position density of SPs.}

\section{Introduction}

Sink pixels (SPs) are pixels that exhibit consistent lower values compared to the background of an image. They likely result from an excess of charge traps localized within individual pixels. They were first studied in the Wide Field Camera 3 UVIS (WFC3/UVIS) channel on the Hubble Space Telescope (HST) \citep{anderson_isr} and \citep{anderson_isr2} and later in the Advanced Camera for Surveys Wide Field Channel (ACS/WFC) \citep{jenna_sink}. Since its installation, HST has been operating in a Low Earth Orbit radiation environment where energetic particles randomly hit the detector, linearly increasing dark current, and creating SPs, and warm and hot pixels. 

The ACS/WFC undergoes a monthly annealing process to address the accumulation of dark current, as well as warm and hot pixels. During the annealing procedure, the WFC CCDs and thermoelectric coolers are powered off, while heaters are activated to raise the CCD temperature from approximately $-81$°C to about 20°C \citep{isr_meaghan}. The purpose of these anneals is to temporarily decrease the dark current and mitigate the occurrence of warm and hot pixels. These cycles do not completely eliminate all warm and hot pixels in the CCDs, leading to a persistent presence of permanent hot pixels in the ACS/WFC over time.

\begin{figure}[H]
  	\centering
 	\begin{subfigure}{1\textwidth}
  		\centering
		\includegraphics[scale=0.49]{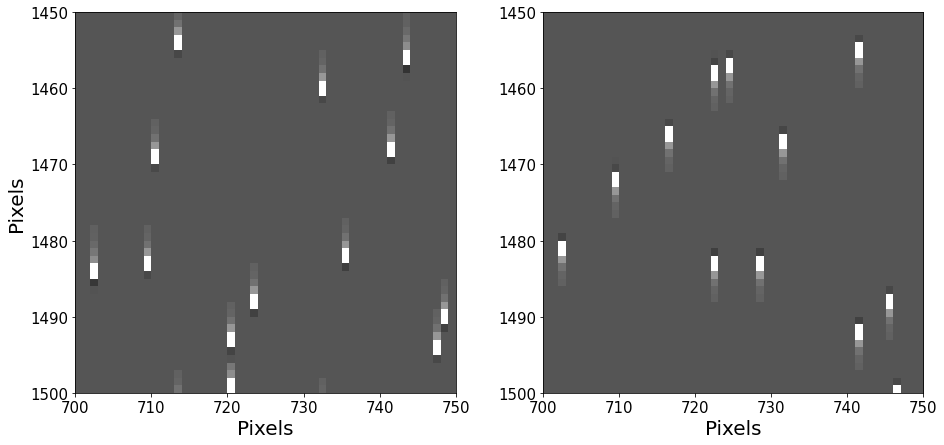}
		\end{subfigure}
 \caption{A 50x50 pixels region of 2021-12-07 \texttt{snk.fits} showing SPs and their trails for WFC1 (left) and WFC2 (right). SPs cause trails in the opposite direction of readout because they trap charge as the data is read out.}
    \label{fig:20211207snk}
\end{figure}

Since 2015, the ACS/WFC has been post-flashing its long (1000.5 sec) and short (0.5 sec) darks each anneal cycle as part of the CCD Daily Monitor Calibration Program. The post-flashed level for all of them is $~60 e^-$ (\texttt{FLASHDUR = 4.6 seconds}). The flashing is performed in an effort to mitigate the impact of charge transfer efficiency (CTE) losses, which are caused by the accumulation of charge traps in the detectors. The dark reference file for each anneal cycle is then constructed by subtracting the average short dark from the average long dark. This removes the post-flash electrons and 0.5 seconds of dark current. The post-flashed short darks revealed a population of low-value pixels, relative to the $~60e^-$ background, that were found to be SPs in the ACS/WFC detector. These pixels are spread randomly across the detector while also affecting the surrounding pixels and creating negative trails in the opposite direction from readout \citep{jenna_sink}. This revelation led to a new reference file, created by the ACS/WFC reference file pipeline, that is used to flag SPs for each anneal cycle since January 2015. These SP reference files are known as \texttt{SNKCFILE} in the Calibration Reference Data System (CRDS), and their filename suffix is \texttt{$*\_$snk.fits}. The application of the SP reference file to science data revealed that approximately $1-2 \% $ of the pixels in the detector were identified as SPs or affected by a SP, and this percentage fluctuates depending on the background level of each science frame \citep{jenna_sink}. Figure \ref{fig:20211207snk} shows a 50 x 50 pixels region of a \texttt{$*\_$snk.fits} file from the 2021-12-07 anneal where the trails caused by SPs can be seen in the detector. 

In this report we'll be focusing solely on SPs and exploring their evolution in the ACS/WFC by analyzing darks and SPs reference files. We will calculate the creation rate of SPs in the detector and assess how the annealing process impacts SPs. We'll analyze how CTE losses affect SP distribution across the detector and create a simulation of short darks to compare the data prior to readout.

\section{Data and Analysis}\label{s:data}

In order to understand the long-term behavior of SPs in the ACS/WFC, we used the SP reference files of the following anneals, 2015-01-15 (CAL/ACS 13952, PI: David Golimowski), 2018-06-26 (CAL/ACS 14948, PI: Tyler Desjardins) and 2021-12-07 (CAL/ACS 16522, PI: Meaghan McDonald) and the post-flashed short darks taken from 2015 through 2021\footnote{Program IDs: CAL/ACS 13952-13954, CAL/ACS 14395-14397,  CAL/ACS 14506, CAL/ACS 14517-14518, CAL/ACS 14946-14948, CAL/ACS 15519-15521, CAL/ACS 15757-15759, CAL/ACS 16377-16379, CAL/ACS 16522}. In the short darks \texttt{blv$\_$tmp.fits} files' header, we set \texttt{FLSHCORR=PERFORM} and \texttt{FLASHDUR=4.6} before processing them through \textbf{acs2d} in CALACS for flash-correction only, where it returned \texttt{flt.fits} files. The SP reference files take into account the full frame including overscans and prescans (4114 x 2068 $\mathrm{pix} ^2$) whereas the short darks consist only of the science frame (4096 x 2048 $\mathrm{pix} ^2$). We then confined the analysis of the SP reference files to exclude physical pre-scans and virtual overscans, focusing solely on the science frame, therefore making it the same dimensions as the \texttt{flt.fits} short dark files.

\subsection{Evolution of Sink Pixels}\label{s:evolution}

Our initial goal was to explore the creation rate of SPs within the detector. We used the most recent SP reference file, from the 2021-12-07 anneal, as a guide to localize SPs. The SP reference file defines anything  $\lesssim$ $-7 e^-$ as a SP, so to be sure we are selecting only bona fide SPs, not noise excursions, we lowered that threshold to $\le -10e^-$. We found 21,168 SPs for WFC1 and 22,900 SPs for WFC2. We then applied the locations of these SPs to the 2021-12-07 \texttt{flt.fits} short dark file, ensuring that our analysis specifically targeted SPs rather than the entirety of the CCD frame. We used this \texttt{flt.fits} short dark file and followed the localized SPs back in time in an attempt to identify when they became SPs. All files from 2015 to 2021 were systematically processed as follows. For each anneal date, we took the set of pixel values prior to the date and the set of pixel values after the date from the short dark \texttt{flt.fits}, and calculated the mean and standard deviation of each set. We define a created SP as the difference in the mean pixel value before and after an anneal date that has to be at least five times the typical variability of that pixel, or a five sigma deviation. 

To further expand on this analysis, we decided to investigate whether SPs can ``heal" or return to average pixel values with each annealing process. We used the same SP threshold of  $\le -10 e^-$ to acquire the locations of these pixels in the oldest SP reference file, from the 2015-01-15 anneal, and found 16,845 SPs for WFC1 and 18,206 SPs for WFC2. The locations of these SPs were then applied to the 2015-01-15 \texttt{flt.fits} short dark file. This \texttt{flt.fits} file was used for reference while we processed the 2015 through 2021 \texttt{flt.fits} short dark files, but this time comparing each one of the files to its subsequent file. In this case, for a SP to have ``healed" we required the pre-break mean to be negative and the post-break mean to be positive, and the difference between the means to be less than the five times scatter before and after the break.

These statistics were used to assign appropriate flags to individual pixels based on which criteria they met. The flagged date is the anneal date when a pixel became a SP, and stayed as such, or ``healed". These statistics allowed us to create Figures \ref{fig:turnedonsinks} and \ref{fig:turnedoffsinks}, where the SPs are plotted over time, to visually examine how the pixel value changed over the course of seven years. We used this information to calculate the accumulation rate of SPs over time, setting zero as the initial condition, and used \textbf{scipy.stats.linregress} to find a line of best fit for each detector.

\section{Initial Results}\label{s:initialresults}

\begin{figure}[H]
  	\centering
 	\begin{subfigure}{1\textwidth}
  		\centering
		\includegraphics[scale=0.245]{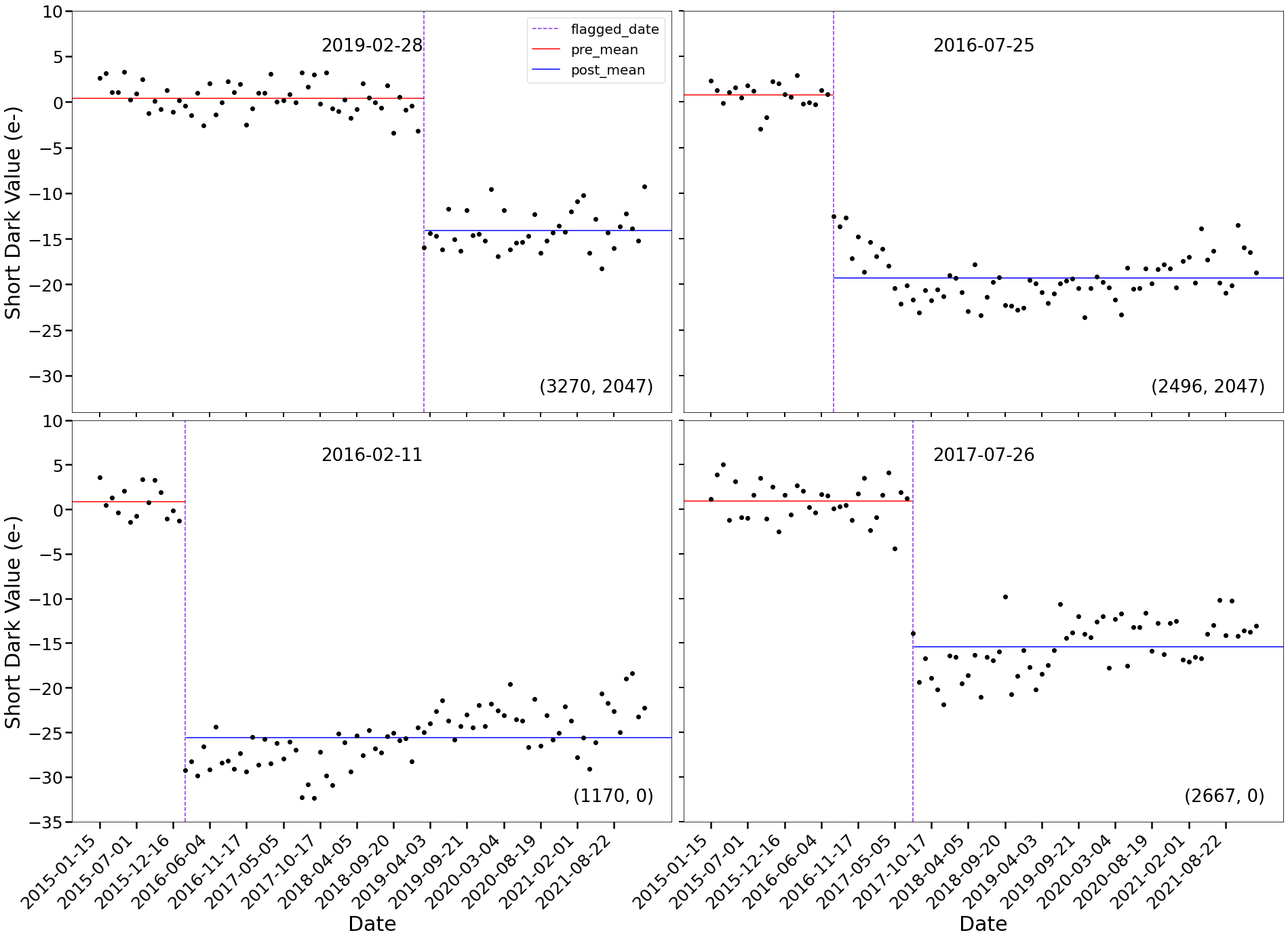}
		\end{subfigure}
 \caption{Four example pixels from WFC1 that were flagged as SPs, along with their creation dates and their locations at the bottom right of each subplot. The red line represents the pre-break mean value of this pixel, before becoming a SP, while the blue line is the post-break mean value. The purple dashed line marks the flagged date of the SP.}
    \label{fig:turnedonsinks}
\end{figure}

We verified that the analysis technique in the preceding section was correctly finding and flagging pixels that became SPs during the seven year period by plotting a few examples as seen on Figure \ref{fig:turnedonsinks}. In Figure \ref{fig:turnedonsinks}, four pixels from WFC1 are plotted to show the behavior of pixels that became SPs. The flagged date is at the top of each graph and the location of the pixel plotted is at the bottom right. The red line represents the mean value prior to the pixel becoming a SP, while the blue line is the mean value once the pixel turned into a SP. The purple dashed line is the flagged date, or creation date, of the SP.  

Some of these SPs have negative values as low as $-47 e^-$. We flagged 5,430 pixels as created SPs for WFC1 and 5,649 created SPs for WFC2. 

\begin{figure}[H]
  	\centering
 	\begin{subfigure}{1\textwidth}
  		\centering
		\includegraphics[scale=0.45]{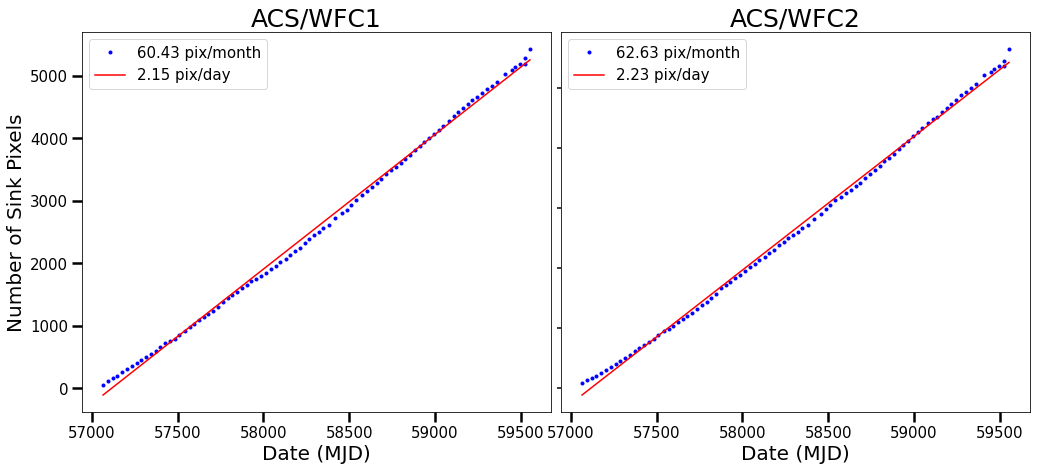}
		\end{subfigure}
 \caption{Accumulation plot of SPs for WFC1 (left) and WFC2 (right). The number of new SPs appearing in each anneal since January 2015 is plotted in blue, while the fitted line is red. There are approximately 60.43 pixels per month that are becoming SPs in WFC1 and approximately 62.63 pixels per month becoming SPs in WFC2.}
    \label{fig:accwfc}
\end{figure}

For WFC1 the creation rate of SPs is 2.15 pixels/day or 60.43 pixels/month while for WFC2 the creation rate is 2.23 pixels/day or 62.63 pixels/month, as seen on Figure \ref{fig:accwfc}. WFC2 has an accumulation rate $ < 1 \% $ higher than WFC1. The data points are shown in blue while the red line is the line of best fit using \textbf{scipy.stats.linregress}. Using this rate we calculated how many SPs are expected to be present in WFC1 for the most recent anneal date. We used the number of SPs in the first 2015 \texttt{flt.fits} short dark as the initial condition and calculated 21,921 SPs expected to be present in the latest file which is very close to the number of SPs $\le -10 e^-$ previously found in the most recent SP reference file: 21,168 SPs. We used the same method to calculate the amount of SPs expected for WFC2, which resulted in 23,466 SPs while the SP reference file contained 22,900 SPs. 

\begin{figure}[H]
  	\centering
 	\begin{subfigure}{1\textwidth}
  		\centering
		\includegraphics[scale=0.245]{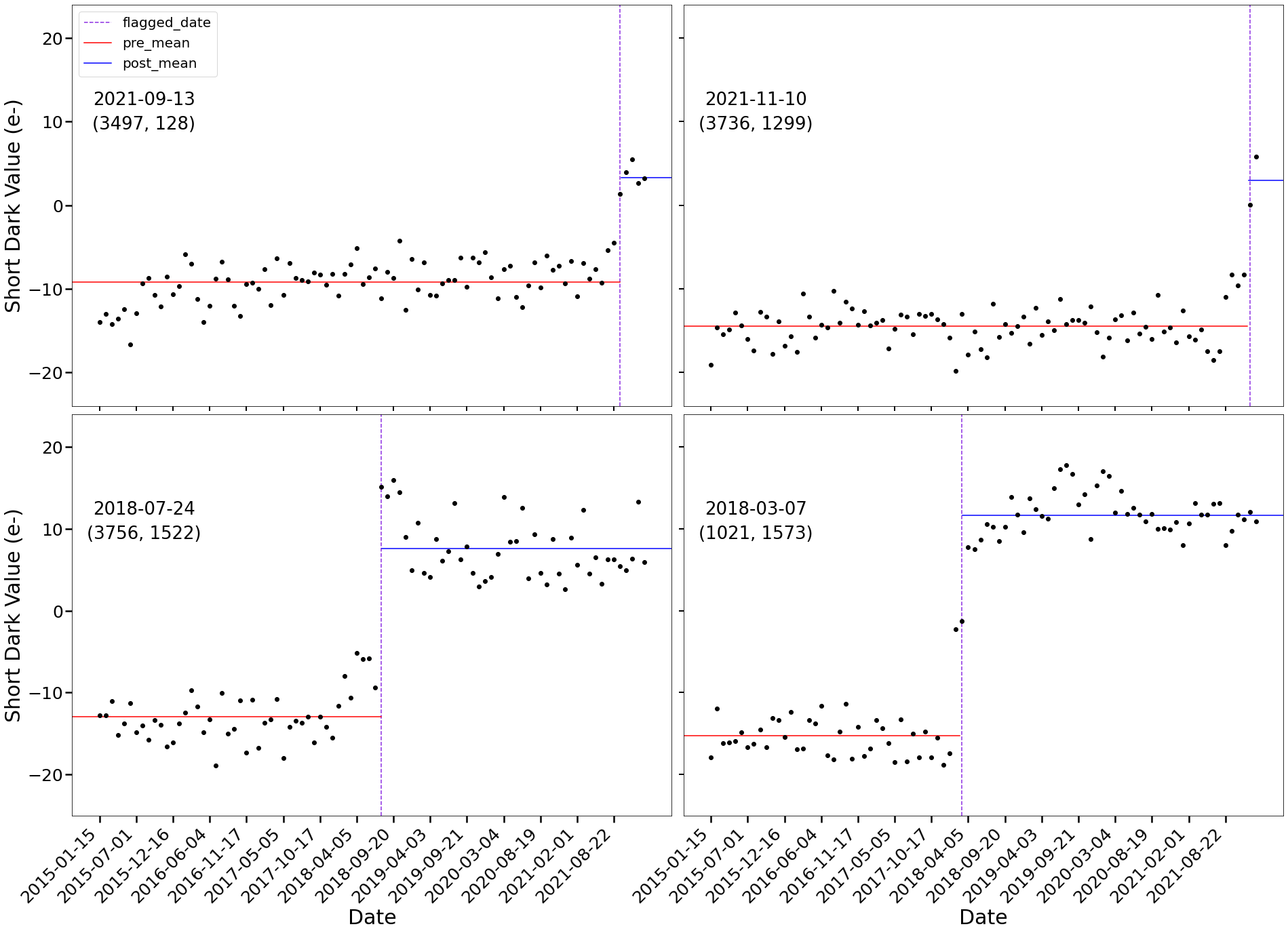}
		\end{subfigure}
 \caption{Four examples out of 11 healed SPs from WFC1 that became regular pixels in the time frame of the data set, along with their flagged dates and their locations at the top left of each subplot. The red line represents the pre-break mean value of this SP, before becoming a regular pixel, while the blue line is the post-break mean value. The purple dashed line marks the flagged date of the healed pixel.}
    \label{fig:turnedoffsinks}
\end{figure}

Since the CCD undergoes monthly anneals to mitigate the occurrence of warm and hot pixels, we decided to investigate whether these anneals also mitigate the number of SPs in the detector. After running the \texttt{flt.fits} short dark data through the code using the oldest  \texttt{flt.fits} short dark (2015-01-15) as the reference for SP locations, the returned table consisted of 11 pixels flagged as ``healed" pixels for WFC1 and 17 for WFC2. Looking at Figure \ref{fig:turnedoffsinks}  some SPs jump to higher values and become warm pixels randomly. These pixels showed a marked increase in pixel value to a normal or warm level, and stayed that way for several anneals. The number of SPs that either ``healed" or became warm pixels is approximately zero making it extremely rare for the monthly annealing process to heal SPs to an average pixel value.
  
\section{Sink Pixels and CTE Effects}\label{s:cte}

\subsection{FLT Short Darks}\label{s:flt}
\begin{figure}[H]
  	\centering
 	\begin{subfigure}{1\textwidth}
  		\centering
		\includegraphics[scale=0.315]{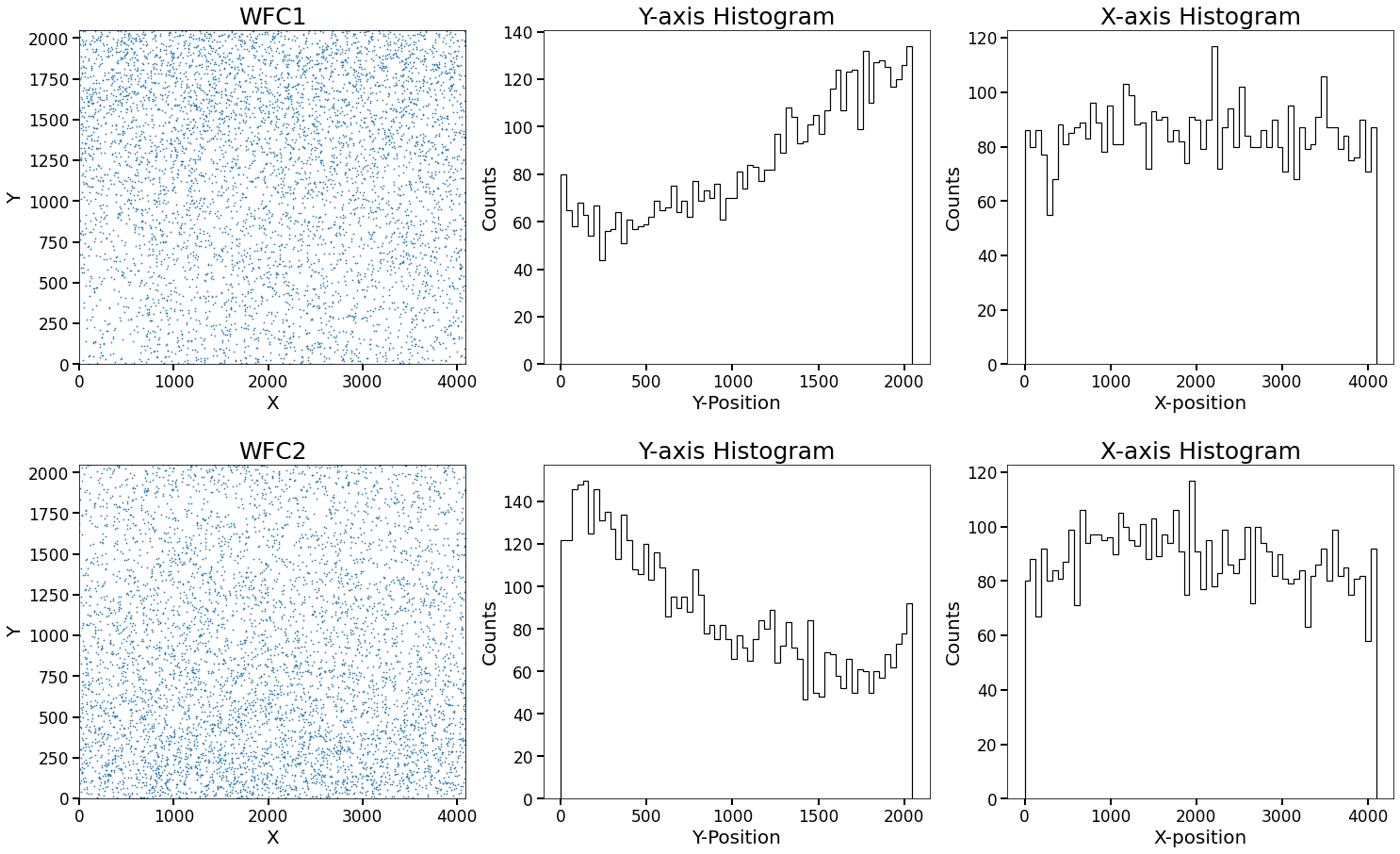}
		\end{subfigure}
 \caption{Locations of pixels flagged as SPs in WFC1 (top) and WFC2 (bottom) science frame from the flash-subtracted short darks (left column), using 2021-12-07 as the reference anneal. The two histograms are the distribution of SPs in the $y$ (middle column) and $x$ (right column) direction. The $y$-pixel range goes from 0-2047 and the $x$-pixel range is 0-4095.}
    \label{fig:histfltwfc}
\end{figure}

The next topic we explored was the spatial distribution of only the newly-created SPs in each chip. We expect to see a random distribution of SPs across the detectors because the radiation damage that creates SPs should be isotropic. We continued to use the \texttt{flt.fits} short darks and plotted the locations of the SPs that were created over 2015-2021, as seen in the leftmost panels of Figure \ref{fig:histfltwfc}. We plot the distribution of $y$ and $x$-positions of these SPs in the middle and rightmost panels. In the WFC1 $y$-position distribution plot we see $40\%$ fewer SPs near the chip gap as compared to near the serial register, with a local peak in the distribution of SPs at the opposite end of the amplifier. For the WFC2 $y$-position distribution, there are approximately $43.3 \%$ more SPs near the serial register. The $x$-position distribution is more uniformly distributed as shown. Similar trends were seen in WFC2, with more SPs being found closer to the readout amplifiers and the $x$ locations being uniformly distributed. Our initial guess was that the post-flash background was affecting the number of SPs closer to the middle of the detector, but because there are no trends seen on the $x$-direction, we continued to investigate the SPs' distribution.

\subsection{SP Reference Files}

\begin{figure}[H]
  	\centering
 	\begin{subfigure}{1\textwidth}
  		\centering
		\includegraphics[scale=0.315]{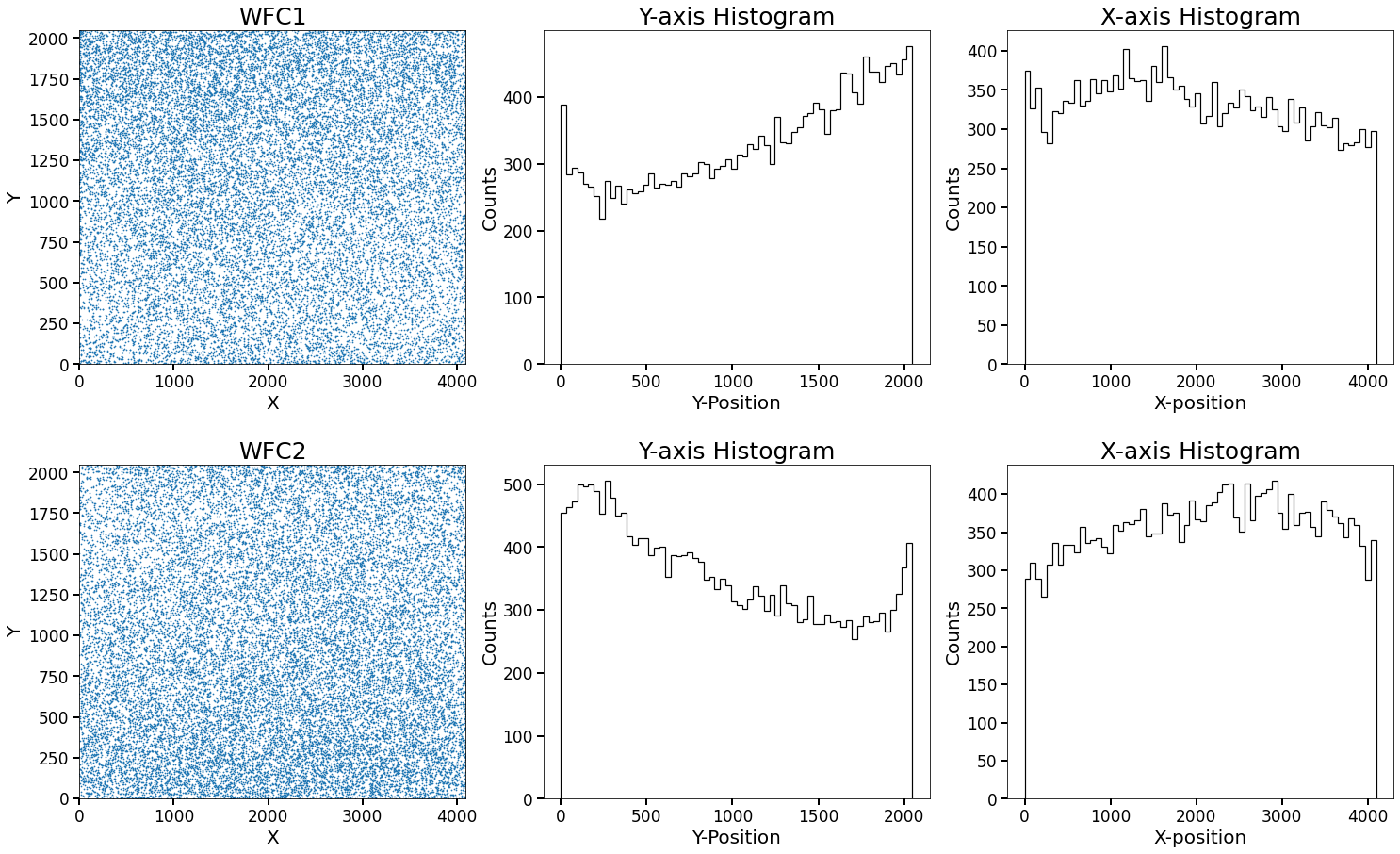}
		\end{subfigure}
 \caption{Same as Figure \ref{fig:histfltwfc} but for the 2021-12-07 SP reference file.}
    \label{fig:histsnkwfc}
\end{figure}

We used the most recent SP reference file to determine whether the total number of SPs would have a random distribution across the detector when plotted. Instead of plotting only the 5,430 (WFC1) and 5,649 (WFC2 ) SPs that were created in these seven years, we chose to look at all the SPs $ \le -10e^-$ from the 2021-12-07 SP reference file. The locations and $x$ and $y$-distributions for the 21,168 SPs in WFC1 and the 22,900 SPs in WFC2 are plotted in Figure \ref{fig:histsnkwfc}. The plots exhibit similar features to those observed in the \texttt{flt.fits} short dark files, such as the presence of a local peak in the distribution of SPs at the opposite end of the amplifier and a gradient in the $y$-position distribution.

In this SP reference file, there is a gradient of the locations of the SPs with approximately $27.5\%$ less SPs detected further away from the serial register for WFC1 and about $20 \%$ less SPs detected away from the readout amplifiers for WFC2. The $x$-position distribution differs slightly with a small left skew for WFC1 and a small right skew for WFC2.

\subsection{FLC Short Darks}

\begin{figure}[H]
  	\centering
 	\begin{subfigure}{1\textwidth}
  		\centering
		\includegraphics[scale=0.315]{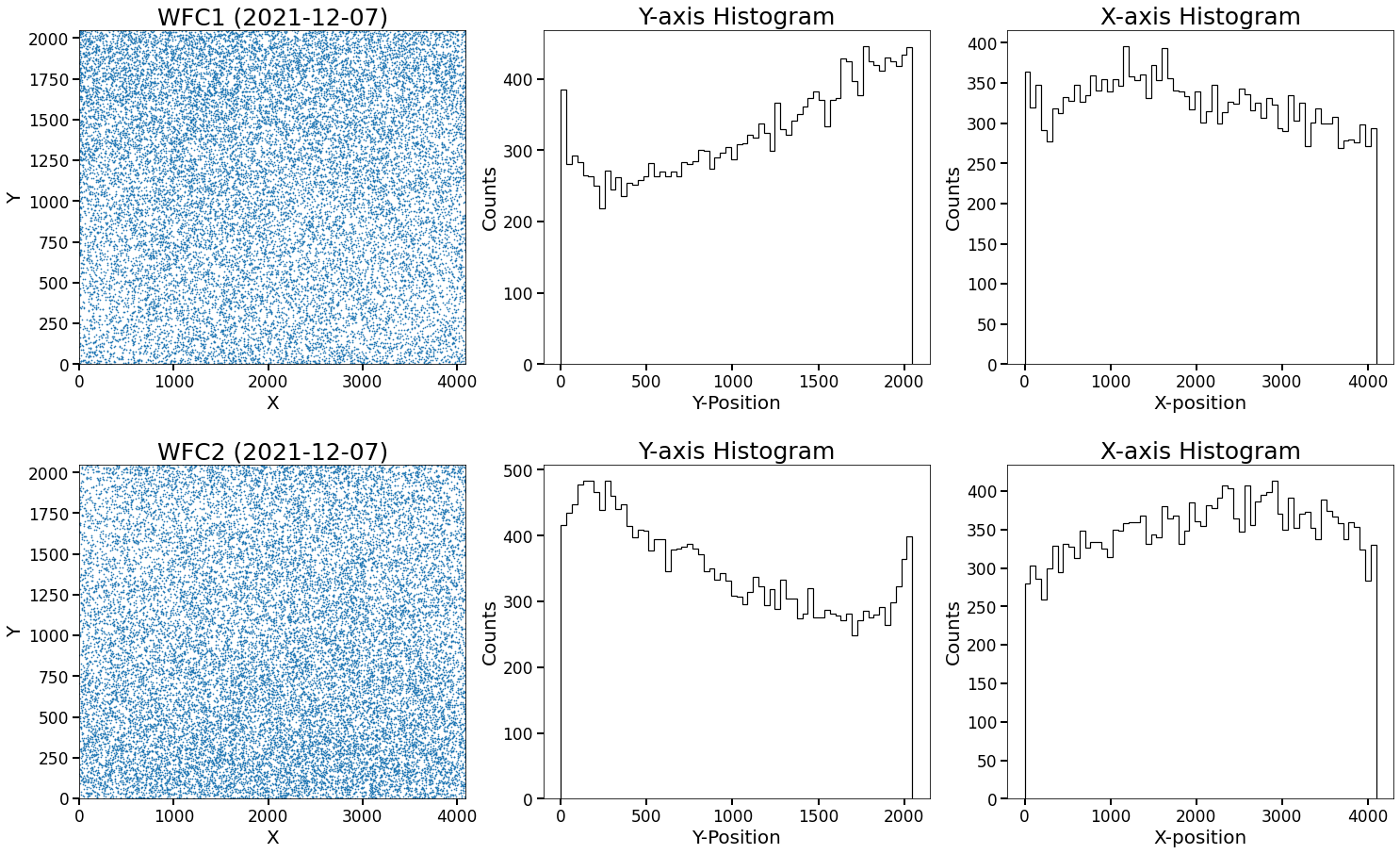}
		\end{subfigure}
 \caption{Same as Figure \ref{fig:histfltwfc} but for the 2021-12-07 CTE-corrected short dark \texttt{flc.fits} file.}
    \label{fig:2021flc}
\end{figure}

\begin{figure}[H]
  	\centering
 	\begin{subfigure}{1\textwidth}
  		\centering
		\includegraphics[scale=0.315]{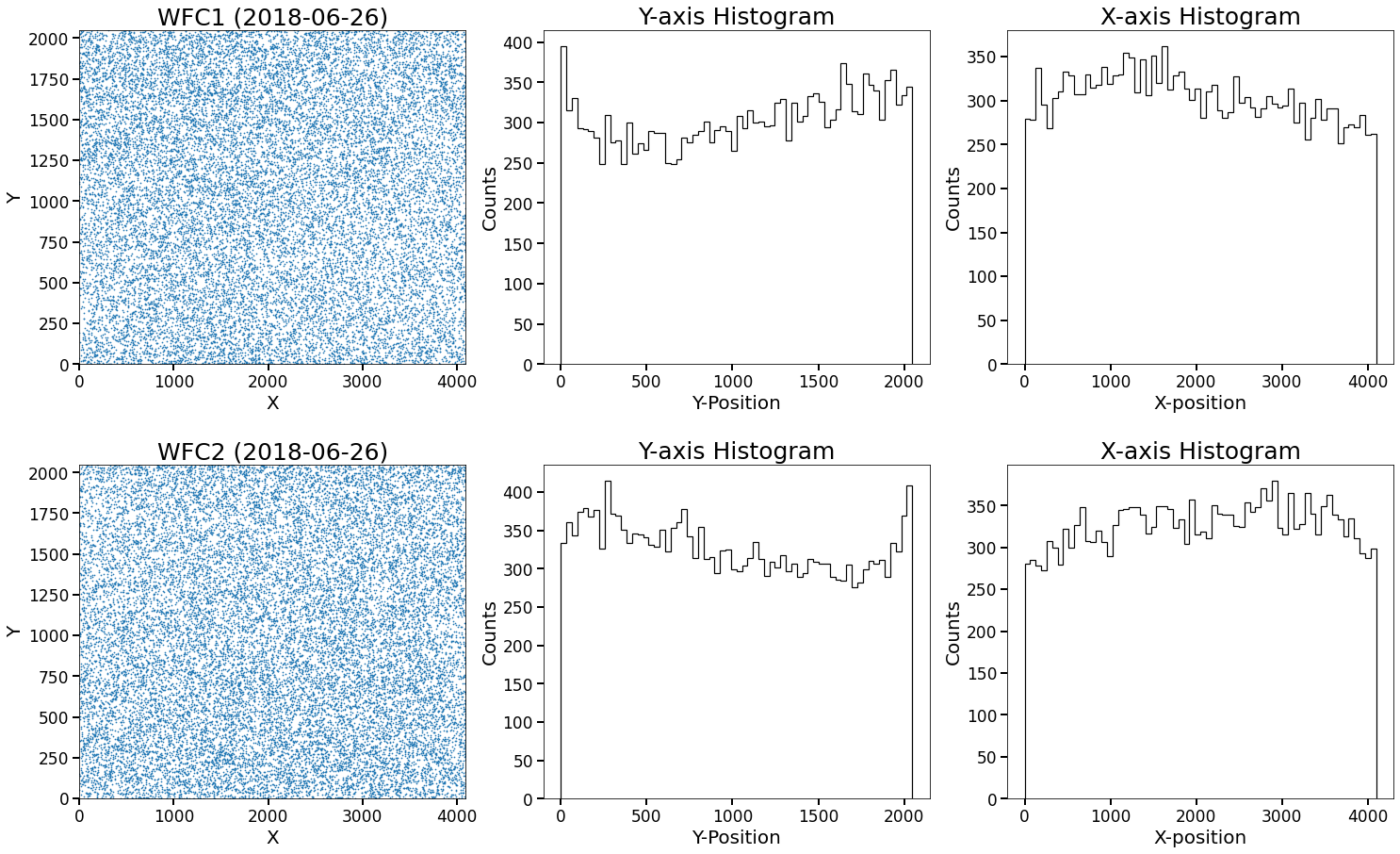}
		\end{subfigure}
 \caption{Same as Figure \ref{fig:histfltwfc} but for the 2018-06-26 CTE-corrected short dark \texttt{flc.fits} file.}
    \label{fig:2018flc}
\end{figure}

\begin{figure}[H]
  	\centering
 	\begin{subfigure}{1\textwidth}
  		\centering
		\includegraphics[scale=0.315]{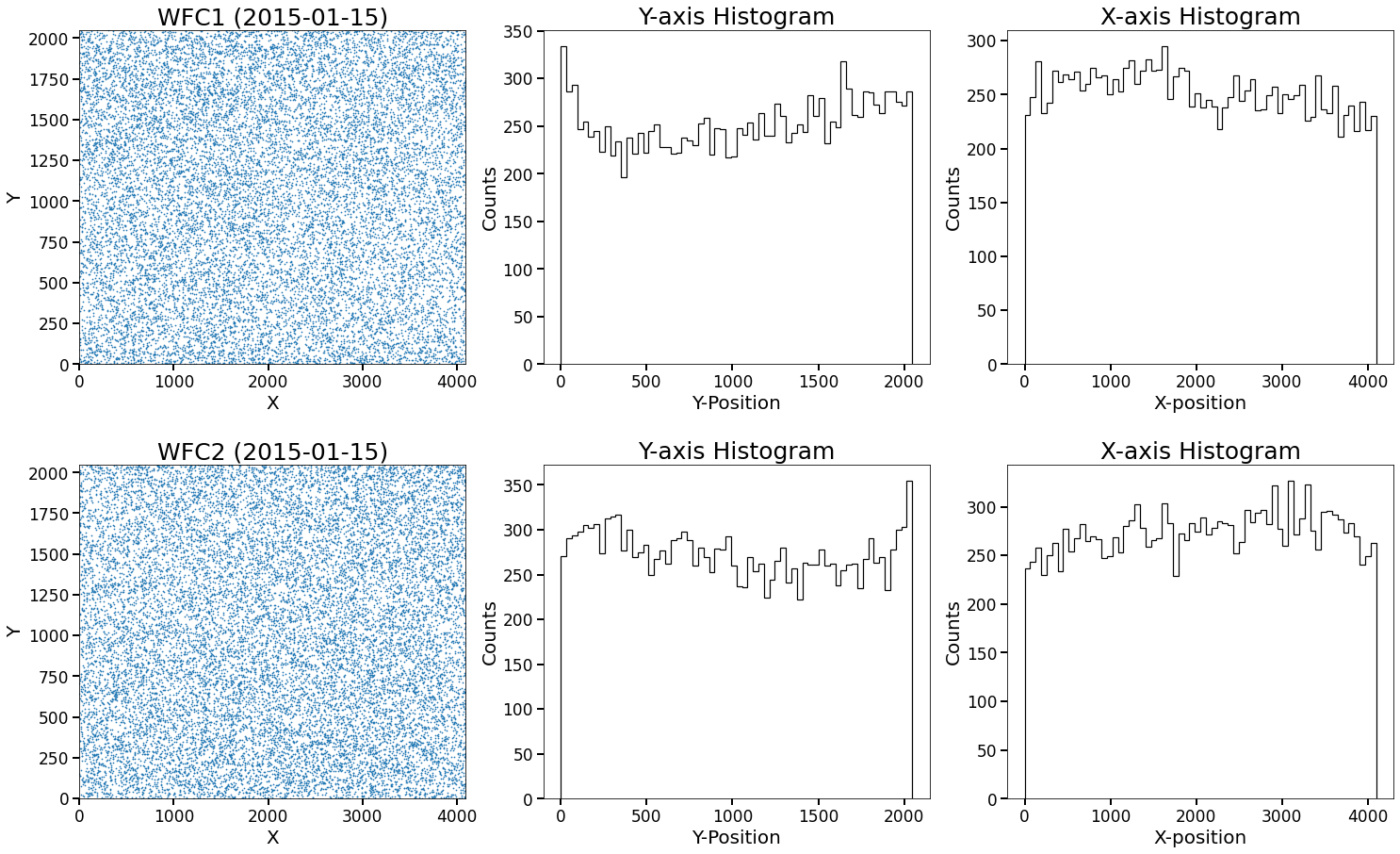}
		\end{subfigure}
 \caption{Same as Figure \ref{fig:histfltwfc} but for the 2015-01-15 CTE-corrected short dark \texttt{flc.fits} file.}
    \label{fig:2015flc}
\end{figure}

Our next step was to assess whether correcting for parallel CTE losses would impact the spatial distribution of SPs in the $y$-direction, specifically examining whether the CTE correction would address the gradient observed in both the SP reference file and the \texttt{flt.fits} short dark file. We begin with the \texttt{blc\_tmp.fits} short dark files, which are CTE-corrected images, from the three different anneal dates, 2021-12-07 (Figure \ref{fig:2021flc}), 2018-06-26 (Figure \ref{fig:2018flc}), and 2015-01-15 (Figure \ref{fig:2015flc}), to see whether the distribution shape changes over time. We performed flash correction on the three files with \textbf{acs2d} resulting in \texttt{flc.fits} files.

The plots in Figure \ref{fig:2021flc} are very similar to that of Figure \ref{fig:histsnkwfc}. The skewed effect is still seen regardless of correcting for CTE. On 2021-12-07, there were approximately $33\%$ fewer pixels near the chip gap for WFC1 and $29\%$ fewer pixels near the chip gap for WFC2. Looking at data from three years prior in Figure \ref{fig:2018flc}, the overall shape of the $y$-position distributions is similar, but the gradient is weaker for earlier data. For WFC1 there were around $19.4\%$ fewer SPs near the chip gap and for WFC2 there were around $17.9 \%$ fewer SPs detected near the chip gap. This may be due to fewer SPs present in the detector during this time. The local peak in the distribution of SPs at the opposite end of the amplifier is greater in this year and it surpasses the count of SPs detected near the chip gap than closer to the readout amplifiers. 

This is more evident in the first \texttt{flc.fits} short dark taken on 2015-01-15, where for WFC1 $15 \%$ less SPs were detected by the chip gap and in WFC2 it was approximately $13\%$ less SPs.  In Figure \ref{fig:2015flc}, because there are fewer SPs detected, the distribution looks more uniform in the $y$-direction but, for WFC1, the first 64 rows, furthest away from serial register, are higher in SP count than the last 64 rows, closer to the serial register. We see the same effect for WFC2. The SP distribution in the $x$-direction showed little to no change throughout the years.

\section{Simulation of Short Darks and Further Results}\label{s:simulation}
To further investigate the reason for the gradient in the distribution of SP $y$-positions, we decided to simulate a short dark image with SPs that are distributed randomly and uniformly across the detector, prior to readout. This will allow us to better understand whether CTE affects SPs when they are being read out.

We begin with the 2021-12-07 CTE-corrected superdark (\texttt{dkc.fits}), which contains the best estimate of the dark rate in each pixel for that anneal period. To simulate a short dark, we first set all negative dark rates to zero, then multiplied the dark rates by the \texttt{DARKTIME} header value from the raw 2021-12-07 short dark. We then Poisson resampled the dark current in each pixel. We used the previous post-break mean values for the SP values in this sample and randomly scattered them uniformly across the detector. Then we multiplied the 2021-12-07 flash file (\texttt{fls.fits}) data by the \texttt{FLASHDUR} (4.6 sec), poisson resampled the flash in each pixel, and lastly added the resampled flash to our simulated short dark.

\begin{figure}[H]
  	\centering
 	\begin{subfigure}{1\textwidth}
  		\centering
		\includegraphics[scale=0.315]{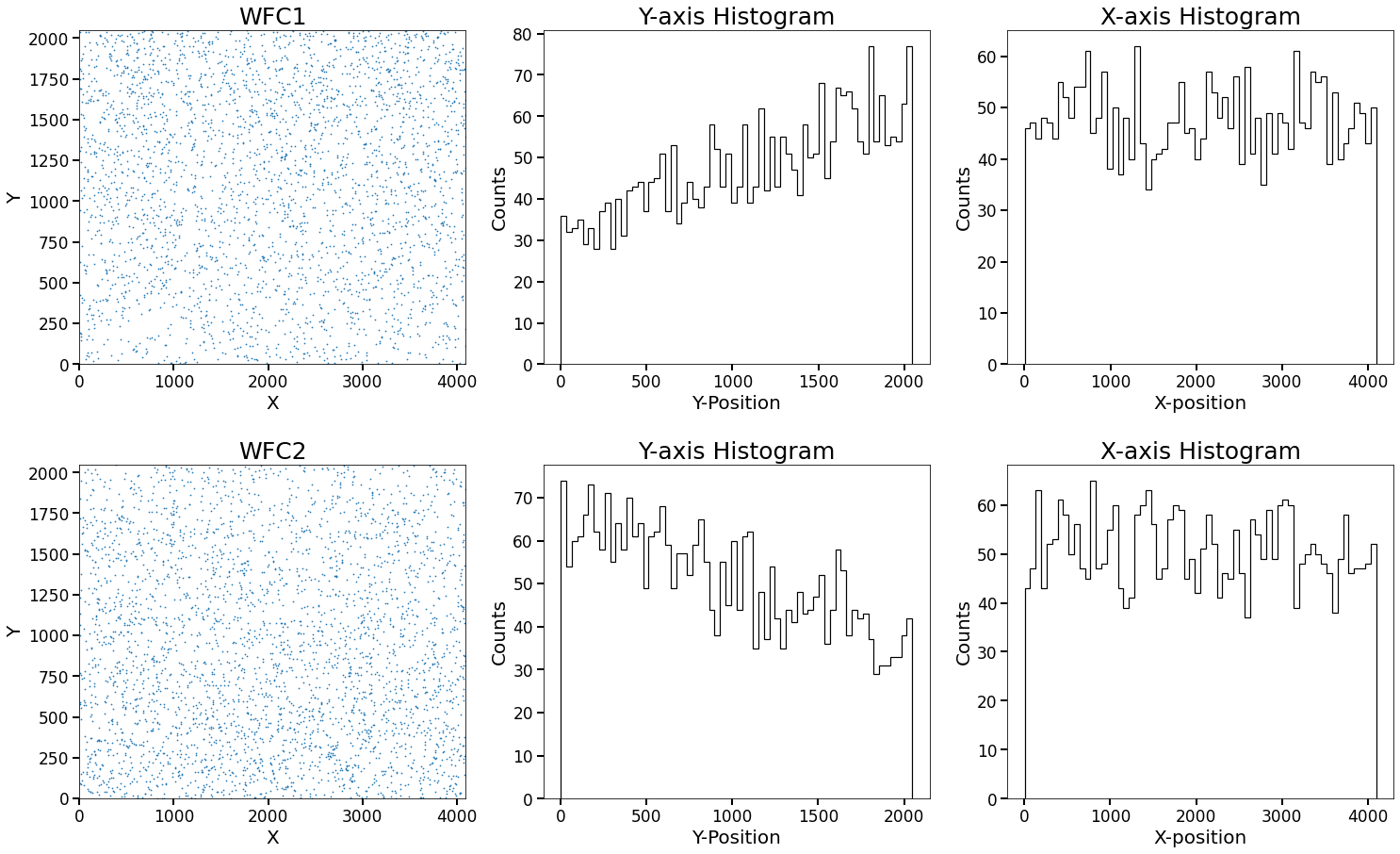}
		\end{subfigure}
 \caption{Simulated WFC1 (top) and WFC2 (bottom) flash-corrected short dark data. The left plots are showing where the SPs are located within each detector. The two histograms are the distribution of SPs in the $y$ and $x$ direction. The $y$-pixel range goes from 0-2047 and the $x$-pixel range is 0-4095.}
    \label{fig:histsimulation}
\end{figure}

We ran our simulated data through the \texttt{acscteforwardmodel} within \texttt{acstools} and corrected for flash using \textbf{acs2d}. Our final product consisted of a simulated, flash-corrected short-dark. We looked at the same SPs we distributed across the detector to determine whether CTE losses during readout had an effect on their depths. In Figure \ref{fig:histsimulation} we have the positions and $x$ and $y$-distributions of this simulated data.

We see similar patterns and distributions in both the $x$ and $y$-direction in the detector to those in the short darks and SP reference files. The difference in SPs found closer to the serial register is $57 \%$ for WFC1 and $44\%$ for WFC2. This provides further evidence that SPs further from the serial register are harder to detect than those closer to it, possibly because SP charge traps are preferentially filled in by CTE-trailed electrons further from the serial register. However, the local peak in the distribution of SPs at the opposite end of the amplifier is not visible and it seems that this effect is only seen across the $y$-direction, as the $x$-direction plot shows us a uniform distribution. The cause of the local peak is therefore unclear, and further work is needed to establish its origin.

\section{Conclusion} \label{s:conclusion}

In this report we studied SPs in ACS/WFC using a variety of datasets taken between 2015-2021. SPs appear to be pixels that contain a large number of charge traps. Consequently, when they fill with background during the post-flash, many of these electrons become trapped within the pixel, failing to follow the other electrons associated with that pixel in the parallel shift to the readout register. As such, these pixels read out low compared to their neighbors. For our purposes, we defined SPs to be pixels whose value is $\le-10 e^-$. We used seven years of short dark data to understand how often SPs are created and if they ever return to average pixel value. We restricted our data to use only the science frame and ignore the physical pre-scans and virtual over-scans. We analyzed variability of SP values over time to flag which pixels became SPs throughout 2015-2021 and stayed $\le-10e^-$ during this time. 

There were 5,430 SPs that were created during this seven year period for WFC1 and 5,649 SPs created for WFC2, which implies a SP creation rate of approximately 2.15 pixels/day for WFC1 and 2.23 pixels/day for WFC2. We used the amount of SPs found in the first 2015 \texttt{flt.fits} short dark file as our initial condition, and the creation rate of SPs to calculate the number of SPs expected to be present in the detector in 2021. By the end of 2021, there were approximately 44,068 SP across the ACS/WFC detector, which makes up around $0.25 \%$ of the science frame, not including SP trails. We then examined whether SPs returned to their previous average values over each annealing process and found that it's extremely rare that SPs get ``healed" and some SPs became warm pixels. 

We then studied the spatial distribution of SPs appearing during the time period 2015-2021, and found that more SPs are located near the serial register than the chip gap in both chips. This gradient persisted in the distribution of all SPs in 2021. To get an explanation for these distributions we decided to look at CTE-corrected images, \texttt{flc.fits}, from three different years, 2021, 2018 and 2015. The gradient gets steeper over time which suggests that CTE losses may be causing the gradient during the readout process, since CTE losses get worse over time and are worse for pixels farther from the serial register. The current process to correct for CTE does not fix this gradient but this could due to the nature of SPs and needs further research. 

Finally, we created a short dark simulation to understand whether CTE affects how SPs are being read out. We used the previous post-break mean values for SPs and distributed them randomly across the detector. This resampled data was then processed and corrected for flash. Once we looked at the same positions of where these SPs were randomly located, we plotted their distribution and noticed that the gradient effect was still present but could not recreate the local peak in the distribution of SPs at the opposite end of the amplifier. CTE losses accumulated during the readout process do appear to create the gradient in the $y$-position of SPs, since the data was initialized with a uniform distribution and the result had a gradient in the direction we expected.

\section{Acknowledgements}
We thank Norman Grogin for assistance, feedback and support through-out the project. We also thank the following ACS team members for providing comments to improve this report: Amy Jones, Christopher Clark, David Stark, Gagandeep Anand, Jay Anderson, Meaghan McDonald and Nimish Hathi. 

For this report we used the following: \texttt{jupyter} \citep{jupyter}, \texttt{numpy} and \texttt{scipy} \citep{cite_numpy}, \texttt{pandas} \citep{pandas}, \texttt{astropy} \citep{astropy}, and  \texttt{matplotlib} \citep{matplotlib}.

\bibliographystyle{apj}
\bibliography{guzman_sinkpixels}

\end{document}